\documentclass[10pt,conference]{IEEEtran}
\usepackage{cite}
\usepackage{amsmath,amssymb,amsfonts}
\usepackage{svg}
\usepackage{comment}
\usepackage{hyperref}
\usepackage{setspace}
\usepackage{caption}
\usepackage{subcaption}
\usepackage{graphicx}
\usepackage{xcolor}
\usepackage{soul}

\definecolor{myred}{HTML}{F8CECC} 
\definecolor{myblue}{HTML}{DAE8FC} 
\definecolor{mygreen}{HTML}{D5E8D4} 
\definecolor{myviolet}{HTML}{E1D5E7} 

\usepackage[norelsize, linesnumbered, ruled, lined, boxed, commentsnumbered, noend]{algorithm2e}

\newcommand{\RNum}[1]{\uppercase\expandafter{\romannumeral #1\relax}}

\newlength\myindent
\def\BibTeX{{\rm B\kern-.05em{\sc i\kern-.025em b}\kern-.08em     T\kern-.1667em\lower.7ex\hbox{E}\kern-.125emX}}

\begin{document}
\title{QoS Aware Mixed-Criticality Task Scheduling in Vehicular Edge Cloud System}

\thispagestyle{plain}
\pagestyle{plain}

\author{\IEEEauthorblockN{Suvarthi Sarkar, Aditya Trivedi,  Ritish Bansal, Aryabartta Sahu \textit{IEEE Senior Member}}
\IEEEauthorblockA{\textit{Dept. of CSE,  IIT Guwahati, Assam, India.} Emails:\{s.sarkar, atrivedi, rbansal, asahu\}@iitg.ac.in }
}

\maketitle

\begin{abstract}

Modern-day cars are equipped with numerous cameras and sensors, typically integrated with advanced decision-control systems that enable the vehicle to perceive its surroundings and navigate autonomously. Efficient processing of data from the decision-control units of sensors, lidars, radars and cameras is quite computationally intensive and can not be done with full-proof accuracy using less capable onboard resources of autonomous vehicles (AVs). In order to deal with this problem, some computation requirements (also referred as tasks) are offloaded to infrastructure or executed in parallel in both AV and infrastructure to enhance accuracy. The infrastructure comprises base stations, a centralized cloud, and a centralized scheduler. Base stations (BSs), located closer to AVs, execute tasks in collaboration with a significantly more powerful centralized cloud, while the centralised scheduler centrally schedules all the tasks. The base station receives tasks from multiple AVs, each with varying deadlines, criticality, and locations. Our main goal is to maximize the profit of the infrastructure by (a) maximizing the number of tasks completed/ minimizing the number of drop tasks, (b) minimizing the distance cost for task offloading, and (c) minimizing the energy usage of BSs. 

In this work, we proposed efficient approaches to schedule the collection of tasks to the BSs, by employing a hybrid scheduling approach where tasks from AVs get allocated to nearby base stations if the nearby BSs are lightly loaded, otherwise AVs send the task to centralised scheduler for allocation. The centralized scheduler maximizes the profit by following strategies: (a) selection of BS considering the distance and energy consumption,  (b) when the task load is moderate or low, highly critical tasks run at favourable utilisation, and (c) low-critical tasks are dropped to free up computational resources for executing high-critical tasks. Based on our extensive experiments, both the proposed approaches improved the quality of service provided by up to 25\% compared to the state-of-the-art approach in real-life datasets.

\end{abstract}


\section{Introduction}

The automotive industry is currently experiencing a notable trend in the rapid advancement of autonomous vehicles (AVs) \cite{Raviteja18}. By the term AV, we refer to vehicles with ADAS (Autonomous Driving Assisted System). Although the ultimate goal of achieving fully autonomous vehicles has not yet been realized, substantial strides have been made from conventional manually-operated vehicles. Researchers have classified AVs into five levels, with level five denoting vehicles capable of operating anywhere in the world without human intervention \cite{Kim21}. Presently, the AV industry has reached approximately level three. To replace human drivers, AVs heavily rely on data collected from their surroundings. Modern advanced cars are equipped with abundant data sensors, LiDAR, radar, and cameras. These AVs process the gathered data through computations, making decisions based on the analyzed output. The underlying concept is that, with each step up the automation hierarchy, there is an augmentation in the volume of data gathering and required computation \cite{ku2017}. 

In fact, modern AVs commonly integrate multiple LiDARs, radar systems, and cameras, ranging in quantity from three to fifteen, the variation in number is observed from one vehicle to another \cite{Liu21}. However, the challenge arises when we consider the computational demands of the sensor-generated data of the AVs. Depending on the results of the computation, the AVs make decisions and drive autonomously. Meeting the computation demands requires an on-board computer system with very high processing and storage capacity. Equipping vehicles with computer systems with such computing and storage capacity can significantly inflate the cost of the vehicle, potentially making the AVs unaffordable for the majority of the population. This is clearly an undesirable outcome. To bypass this, the AVs connect themselves with infrastructure and offload a portion of their computational requirements to the infrastructure. The infrastructure essentially provides remote edge computing service to the AVs for low-latency, real-time processing \cite{Dai22}. An edge cloud computing platform is a distributed computing system that processes data closer to the source of the data (this is case, the AVs are source of data), reducing latency and enhancing real-time processing capabilities. In this context, we refer to these computational requirements of AVs as ``tasks".

To effectively manage the decision-making process for task offloading in autonomous vehicles (AVs), tasks are categorized into three categories. 
\begin{itemize}
    \item High-level critical tasks: at the highest criticality level, tasks require minimal latency, with a typical latency window of 10-20 milliseconds. The highest critical tasks include acceleration control, braking, pre-crash sensing warnings, automated overtaking, etc. \cite{Ali20}.
    \item Mid-level critical task: the intermediate tier of critical tasks includes operations like object detection, road safety services, queue warning, etc. These tasks are processed both locally within the AVs and remotely on base stations (BSs) in parallel. Given the limited computational capabilities of AVs, onboard deployment involves lightweight models with somewhat reduced accuracy, complemented by premium models on latency-sensitive edge servers. More examples of tasks in this mid-level category include emergency warnings, cooperative collision avoidance, queue warnings, and various road safety services, which require a latency of approximately 100 milliseconds \cite{Ali20}.
    \item Low-level critical task: at the lowest level of criticality, tasks such as destination path finding, infotainment operations, media functions, speech recognition etc. They are exclusively processed on remote BSs, operating with more relaxed deadlines, necessitating latencies ranging from 500 milliseconds to 5 seconds \cite{Ali20}.

\end{itemize}

In this study, our perspective revolves around infrastructure considerations, focusing on task scheduling to minimize the overall infrastructure costs for task execution. Given that only mid-level and low-level critical tasks are offloaded to the infrastructure, we categorize these tasks directed to infrastructure as ``hard" and ``soft" tasks, respectively. This classification is established based on the severity of penalties associated with missed deadlines. Hard real-time tasks result in large penalties if deadlines are not met, while soft real-time tasks incur lighter penalties in such instances.

We consider a futuristic smart city, which provides computing assistance to all the AVS. We refer to this entire city, along with all the AVs and its computation-providing servers, as a Vehicular Edge Cloud System (VECS) \cite{Wan21, Tang20}. Inside the city, AVs run in abundance. The system has two participants- (a) infrastructure and (b) AVs. 
A group of AVs, establish communication with the infrastructure. The infrastructure acts as a service provider and receives subscription fees from the AVs and, in turn, leases their computational units to meet the AVs' computational demands. The infrastructure consists of base stations (BSs) and cloud system. These BSs are equipped with limited computational capacities while the cloud server offers several orders of magnitude more computing power. The BSs are located geographically much closer to the AVs, and the cloud is connected to all the BSs but located far away geographically. So, the BSs share the computation load among themselves much more often than sharing the computation load with the cloud. To address varying computational workloads, BSs engage in the sharing of their computational load. The BSs periodically communicate their resource availability to the network to facilitate seamless resource sharing, enhancing overall system efficiency. Within this system, AVs assume the role of users who subscribe to the VECS by paying subscription fees. In exchange, they gain access to the computational resources they require, effectively renting infrastructure's resources. The AVs actively generate computational tasks, which they offload to infrastructures for essential computational assistance.

The base station (BS) has two components - (a) Road side unit (RSU), and (b) Edge server (ES). The RSU seamlessly form a connection with each of the AVs in its local area. We consider that there is ample bandwidth to facilitate RSU connectivity with all nearby AVs. The connected AVs offload the tasks to BSs via the respective RSU, and the ES within the BS efficiently manages and fulfils the computational demands generated by these AVs. For the sake of simplicity and focus on our primary objective of delivering good Quality of Service (QoS) through effective computational resource management, we refer to the collection of RSU and ES as BS. This is because of our aim is to address the computational demands by ensuring QoS for the AVs within the system.

Traditionally traffic prediction is possible based on historical data \cite{basepaper}.
However, forecasting the occurrence and frequency of various categories of offloaded tasks presents a considerable challenge. This complexity arises from the fact that the quantity of tasks to offload depends not only on vehicle frequency but also on the diverse habits and activities of vehicle users. Furthermore, predicting the behaviour of AVs adds an additional layer of difficulty, as they share the road with conventional manually-driven cars. This coexistence significantly influences task generation within the AV context. Consequently, traditional offline approaches prove less effective in handling such dynamically generated tasks. In this scenario, tasks arrive in an online fashion, without adhering to periodic patterns, and they come with specific deadlines for completion. 

In this paper, we consider that tasks arrive to the infrastructure with three parameters- execution time, deadline and criticality. The deadline depends on the maximum allowed finish time, and the criticality depends on the importance of the tasks. Here we consider mixed critical tasks. Mixed critical tasks refer to tasks within a system that have varying levels of importance and urgency, requiring different priorities and resources for execution \cite{baruah2}. Mixed-criticality systems are growing more complicated, less uniform and unpredictable.

Our focus is to improve overall QoS efficiently. So we considered the case when various system of BSs works collectively to provide good QoS to the AVs. We view our problem from the infrastructure point of view and try to maximize their profit by (a) maximizing the number of tasks completed/ minimizing the number of drop tasks, (b) minimizing the distance cost for task offloading, (c) minimizing the energy usage of BSs.

The primary contributions of this work can be summarized as follows:
\begin{itemize}
    \item We address the maximization of the total profit of a VECS by minimizing the total cost incurred by the VECS. The total cost consists of task dropping cost, the energy cost of the computation unit, and distance cost due to the distance between the computation unit and autonomous vehicles. Notably, no prior research has integrated three of them together.
    \item We delve into the domain of VECS by focusing on the novel consideration of online mixed-critical tasks with varying deadlines in homogeneous computing environments. To the best of our knowledge, no prior work has explored VECS in this specific context, making our contribution pioneering in this area.
    \item Our proposed methodology demonstrates superior performance when compared to existing state-of-the-art methods across both synthetic and real-life datasets. We conducted extensive experiments across various scenarios, providing detailed explanations for the superior performance of our approach relative to other existing methods.
\end{itemize}
\section{Related Work}

This section reviews existing literature in the domains of vehicular edge cloud computing and real-time task scheduling. 

Tang et al. \cite{Tang20} addressed resource allocation challenges in AVs within real-life scenarios. They considered the AVs connect to one base station, with no communication among base stations. The AVs leave the range of a base station and join another base station. Each base station tries to maximize its own profit. The proposed algorithm, based on backtracking and binary search, aimed to maximize each base station's profit. However, this work solely concentrated on resource allocation, neglecting other factors such as bandwidth availability, base station location, and hands-off considerations.

Dai et al. \cite{Dai22} aimed to minimize service delays requested by AVs in VECS.  Their approach involved connected base stations forming coalitions and sharing tasks based on their respective loads. They designed a probabilistic approach based on historical data to solve the problem. Wan et al. \cite{Wan21} presented a heuristic solution for a modified version of this problem, considering varying computation demands of incoming tasks and introducing a 5G-enabled Edge Computing-Internet of Vehicles (EC-IoV) system framework.  Xu et al. \cite{Xu22} explored the pre-computation of certain tasks by base stations and provided a heuristic solution. It proposes novel strategies to manage computational tasks and cache services, reducing latency and enhancing the overall QoS. Ku et al. \cite{basepaper} aimed to optimize Quality of Service (QoS) in the VECS by considering the energy requirements of base stations to enhance the system's overall profit. The study proposes approaches to manage these resources efficiently, ensuring low latency and high reliability for data-intensive vehicular applications. It addresses the challenges of maintaining consistent service quality despite the dynamic and decentralized nature of vehicular environments. Soni et al. \cite{Soni22} offered an in-depth survey work on computation requirements in vehicular edge cloud systems.
 
A significant portion of the current Internet of Vehicles (IoV) research focuses on 5G technology. Kombate et al. \cite{kombate16} highlighted the advantages of employing a 5G communication model for future IoV implementations, emphasizing low latency, extremely high bandwidth, and reliability. Singh et al. \cite{singh16} conducted a comprehensive review of state-of-the-art research on 5G network technologies.

Notably, many existing works consider factors such as energy requirements, task-dropping penalties, and distance costs. However, to the best of our knowledge, there is a gap in the literature where researchers attempt to maximize the profit of the system by simultaneously minimizing dropping costs, energy costs, and distance costs. Our proposed approach aims to address this gap in the current research landscape.

\section{System Model and Problem Statement}
\subsection{System Architecture}
\begin{figure}
    \centering
    \includegraphics[scale=.88]{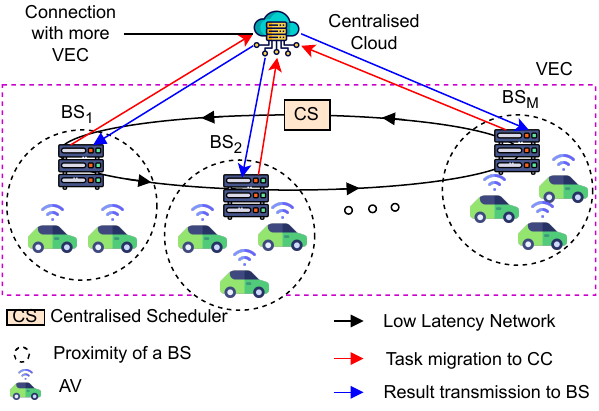}
    \caption{ \footnotesize System Model}
    \label{fig:arc}
\end{figure}

\autoref{fig:arc} depicts the architectural framework of our considered VECS. The considered VECS has a total of $M$ BSs in the VECS infrastructure, which are denoted as $BS_1, BS_2, \cdots, BS_j, \cdots BS_M \in \mathcal{B}$. 

The Centralized Cloud (CC) is connected with all the BSs, creating an interconnected network through a high-speed communication infrastructure that facilitates seamless task migration. The BSs are linked to a Hybrid Centralized Scheduler (CS), which efficiently manages task scheduling, especially during high workload duration of the BSs. During low workload periods, BSs autonomously provide computation services to AVs without notifying the CS. However, when the workload surpasses a predefined threshold, signalling high demand, the CS collects metadata on requested tasks to the infrastructure, assesses the current workload of BSs, and schedules tasks to maximize the profit of infrastructure. The CS selects a BS based on the respective BS's workload and geographical proximity to both the BS and AV to facilitate efficient task offloading. AVs then offload tasks to the selected BS, fulfilling their computation demands. Together, the CC, CS, and BS collectively form the infrastructure and operate with the primary objective of providing computation services to AVs in exchange for monetary benefits from the AVs. 

To provide a concrete illustration, let us consider a scenario where there is a single Centralized Cloud located in a city (for example, New York). The VECS infrastructure covers the entire Manhattan region and the CS operates through the entire city. We assume several BSs distributed across the various city blocks within Manhattan. AVs interact with this infrastructure by connecting to the BS within their local proximity, thus satisfying their computational demands.

Furthermore, we assume that both the BSs and AVs are situated within a single geographical location. Each BS and AV is assigned coordinates, denoted as $(x, y)$, where $x$ and $y$ coordinates represent locations within an $m \times m$ grid.

\subsection{Task Model}

The Autonomous Vehicles (AVs) generate tasks and, based on their internal firmware, determine whether it is necessary to offload these tasks to the infrastructure.  To uniquely identify incoming tasks within the infrastructure, we use an enumeration represented by the variable $i$, and $i^{th}$ task is denoted as $\tau_i$. Let $I$ denote the set of all incoming tasks to the infrastructure and $|I|=N$. Each task is formally defined as by- (a) $id$: identification number of the task. It can be assumed as a combination sequence of number of tasks from an AV and that particular AV with task, (b) $a$: generation time of the task, (c) $d$: deadline of the task, (d) $p$: minimum processing duration of the task, and (e) $flag$: boolean flag, which denotes whether the task is a hard task or soft task.

We classify the tasks into two categories: (a) hard real-time tasks: tasks possess a high level of criticality, requiring the system to prioritize their completion as promptly as possible, (b) soft real-time tasks: in contrast, these tasks have lower criticality levels and are assigned a lower priority compared to hard real-time tasks. Due to the differing penalties for missing deadlines for these two types of tasks, the system is considered mixed critical.

\begin{figure}[tb!]
    \centering
    \includegraphics[scale=.91]{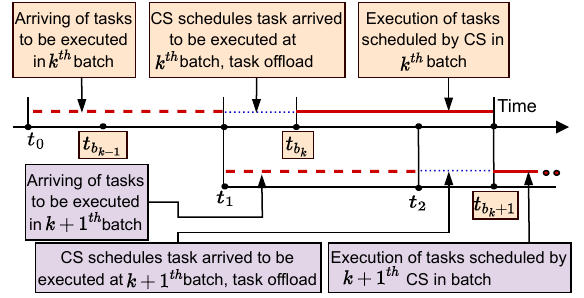}
    \caption{ \footnotesize Order of task arrival, CS scheduling, task migration and execution of tasks in batches}
    \label{fig:time}
\end{figure}

\autoref{fig:time} shows systematic view of order of task scheduling, CS scheduling, task migration and execution of tasks in batches. The tasks that arrive at different BSs between time window $t_0$ to $t_1$ are scheduled by CS followed by task offloading from AV to the respective BS during time window $t_1$ to $t_{b_{k+1}}$, and execute from time window $t_{b_{k}}$ to $t_{b_{k+1}}$. Similarly, tasks arriving between $t_1$ and $t_2$ are scheduled by the CS, and task offloading occurs between $t_2$ and $t_{b_{k+1}}$, with execution starting from $t_{b_{k+1}}$. This systematic view is a key component for understanding the presented task scheduling framework.

Task execution initiates at the beginning of each batch.  Let $s_i$ denote the start time of $\tau_i$. The batch time chosen is small, to keep the model realistic. The interval between two batches is denoted by $t_{\beta}$ units. Consequently, $t_{b_{k}}-t_{b_{k-1}} = t_{b_{k+1}}-t_{b_{k}} = t_{\beta}$. Moreover, we express $s_i = a_i+ \Delta$, such that $s_i = c * t_{\beta}$ and $\Delta \le t_{\beta}$, where $c$ is a positive constant and $\Delta$ represents the delay taken by the infrastructure for task scheduling and migration. $I_{b_k}$ denotes the set of tasks that arrived for execution during $k^{th}$ batch and $\|I_{b_k}\|$ is the total number of tasks submitted by all the AVs in the VECS at the $k^{th}$ batch.

\subsection{Utilisation Model}
\label{um}
We use the term utilisation for both the BSs and the tasks. For clarity, we denote the utilisation of the BSs with \( U \) (uppercase), while the utilisation associated with tasks is represented by \( u \) (lowercase).
Each task utilises the computation resource of a BS by some amount. Due to limited resources, each BS can only process a limited number of tasks based on their resource requirement.

Each task $\tau_i$ must execute for at least $p_i$ time units, having the resource utilisation of $u^{max}$ per unit time. We define $u^{max}$ as maximum amount of CPU resource of BS any task can utilize per unit time, it is constant and equal for all the BSs in the infrastructure.
The processing speed of each BS is adaptable and varies depending on how much utility is allocated by a particular BS to the task \cite{buttazzo2011hard}. So, $\tau_i$ can also execute for some duration $p'_i \geq p_i$, utilizing $u_i$ amount of CPU resource. In that case, $\tau_i$ utilises the CPU of the BS by a lesser amount. Moreover, $p'_i \leq d_i - s_i$, as a task, cannot execute beyond its deadline. Thus, the utilisation is given by 
\begin{equation} 
u_i = \frac{p'_i}{d_i - s_i}
\label{eqn:u_i}
\end{equation}
It is important to highlight that for task $\tau_i$, the values of $u_i$, $d_i$, and $s_i$ are fixed, which may result in $p'_i$ being a fractional value. To simplify our calculations, we adopt the practice of considering $\lceil p'_i \rceil$ as the effective processing time to keep the processing time in a discrete domain.  

Additionally, $u_i \geq u_i^{min}$, where $u_i^{min}$ means the minimum resource any task can utilize per unit of time.
\begin{equation}
\label{eqn:u_min}
u_i^{min} =  \frac{p_i}{d_i - s_i} 
\end{equation}

We use $U_j(t)$ to denote the utility of $BS_j$ used by all tasks at time $t$. $U_j(t)$ falls in the range $[0, U^{max}]$, where $U^{max}$ is maximum usable utility for $BS_j$. The BSs are homogenous, so $U^{max}$ is constant for all the BSs. 

We consider that multiple tasks can execute on the same BS sharing its computation resource with dynamic utilisation depending on the availability of CPU resources of that BS. We define indicator variable $x_{i,j}(t)$ to signify the execution of $\tau_i$ at $BS_j$ at time $t$. The value of $x_{i,j}(t)=1$, if $BS_j$ is executing $\tau_i$ at time $t$, it is $0$ otherwise. The utilisation consumed by all the tasks in a BS should be within its limit all the time and is formally written as:
\begin{equation}
    \label{eqn:utility}
    \sum_{i=1}^N \sum_{j=1}^M x_{i,j}(t) * u_i = U_j(t)\leq U^{max} \:\:\:\forall t
\end{equation}

\begin{itemize}
    \item {The term $U_j (t) $ and $U^{max}$ are associated with $BS_j$. $U^{max}$ is constant for all the BSs in the VECS and typical values of $U^{max}$ are 32, 64, etc. It denotes the total number of computation cores (computational resources) the BS possesses. $U_j (t)$ denotes the number of computational cores that all the tasks executing in $BS_j$ utilize simultaneously at time $t$.}

    \item {The terms $u^{max}$, $u_i$, and $u^{min}_i$ are associated with $\tau_i$. They denote the proportion of a computational core that the task engages, with all three terms normalized to 1. Specifically, $u_i$ and $u^{min}_i$ depend on $\tau_i$ and are calculated using \autoref{eqn:u_i} and \autoref{eqn:u_min}, respectively. The term $u^{max}$ is a constant, and $u^{max} = 1$, which signifies that at max, a task can utilize one full computing core of a BS. Additionally, it satisfies the condition $u^{max} \geq u_i \geq u^{min}_i$.
}
\end{itemize}

\subsection{Energy Model}
A significant amount of energy is consumed by the BS when they execute the tasks generated by AVs. We use the energy model defined by Dayarathna et al. in \cite{paper10}. Energy used by all the BS is denoted by $C_{e}$, which is given by

\begin{equation} 
    \ P_{j}(t) =  P_{s}+ \left[ \left( \left( P^{max} - P_S \right)  + (U_{j}(t)/U^{max})^{3} \right) \right]
\label{eqn:energy}
\end{equation}

\begin{equation} 
    C_{e} = \sum_{\forall t \in \mathcal{T}}  \sum_{j=1}^M P_j(t)
\end{equation}

Here $P_S$ represents the static power consumption required by a $BS$. $P_{j}(t)$ represents the amount of electric power used by $BS_j$ at time $t$, $P^{max}$ represents the maximum eclectic power consumption by BS. \textcolor{black}{For example,  suppose the BS have 64 cores, the maximum utilisation capacity of the BS equals to 64 units ($U^{max}$), and again consider, the electric power requirement ($P^{max}$) is 2000 units to use all the cores at full capacity. If at a given time $t$, the available solar energy is 1500 units ($P_{j}(t)$), then the ES can use approximately 56 CPU cores ($U_j(t)$).}

We define $\lambda_e$ as cost per energy unit, so $\lambda_e * C_e$ denotes the total cost the infrastructure pays for all its energy usage.

\subsection{Distance Model}
We consider that all the BSs and AVs are located in a small geographical region, say a city. We consider the city to be $m \times m$ grid. In this grid, the integral coordinates are valid locations for the BSs and AVs. There are certain fixed positions where BSs are located. \textcolor{black}{AV requests to connect to the VECS infrastructure. After receiving requests, the CS chooses a BS for the AV to connect to and offload tasks. So, the task offloading cost of infrastructure is given by $C_{dis}$ at time $t$.}

\begin{equation} 
    \ C_{dis} =   \sum_{i=1}^N \sum_{j=1}^M {d_{i,j}} * y_{i,j}
\label{eqn:dis}
\end{equation}

Here $d_{i,j}$ is calculated as $D(Loc(AV_l),\: Loc(BS_j))$,
where D is Euclidean distance, $Loc(AV_l)$ denotes location of $i^{th}$ task generating AV (let us consider $AV_l$ generates $\tau_i$) and $Loc(BS_j)$ denotes location of $j^{th}$ BS (on which task is scheduled). We define indicator variable $y_{i,j}$, the value of $y_{i,j}=1$ if $x_{i,j}(t)=1$ for at least one value of $t$, in range $s_i \leq t \leq s_i+p'_i$, $0$ otherwise. In other words, $y_{i,j}=1$ if $\tau_i$ is scheduled to $BS_j$ for execution. 

$C_{dis}$ denotes the total distance cost of the infrastructure. We define $LOC_{BS}$ as a vector to store the location of all the BSs. The vector $LOC_{AV}^{t_k}$ denote the location of all AVs at the start of $t_{b_k}$ batch. The vector $LOC_{BS}$ is constant, as the location of BS does not change, but the location of AVs changes with time, so the vector $LOC_{AV}^{t_k}$ is dependent on time.

In preparation for task execution, AV offloads their computational request (task) to the BS that offers computational services. The task data passes through various intermediate BSs to reach the intended BS from the AV, similarly the intended response travel the opposite way round. The number of hops through intermediate BSs increases with greater distance, increasing the distance cost. Let $\lambda_{dis}$ denote the cost the infrastructure pays per unit distance. So, $\lambda_{dis} * C_{dis}$ denotes the total cost the infrastructure pays for task offloading.

\subsection{QoS Model}

The assessment of Quality of Service (QoS) is determined based on the occurrences of service outages within the workloads. We modified the QoS model used by Dey et al. in \cite{basepaper}. Given that both energy and computing resources are constrained at BSs, it is possible that BSs may not have the capacity to fulfil all the computational demands from AVs while still meeting the specified deadlines. 

A service outage occurs when its workload cannot be executed by the VECS infrastructure, hence can not be completed within its deadline. We define an indicator variable $z_{i,j}$, which denotes the dropping of $\tau_i$ from $BS_j$. So $z_{i,j}=0$ if $x_{i,j}(t)=1 \:\forall t$ in range $s_i \leq t \leq s_i+p'_i$, $1$ otherwise.

\begin{figure}
    \centering
    \includegraphics[scale=.71]{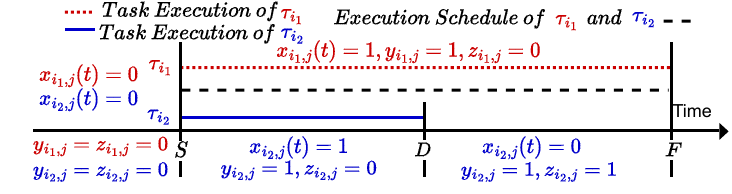}
    \caption{ \footnotesize Tasks $\tau_{i_1}$ and $\tau_{i_2}$ is scheduled in $BS_j$ from time S and F. But at time $D$ task $\tau_{i_2}$ is dropped. The figure shows how the value of variables $x_{i,j}(t),\:y_{i,j}$ and $z_{i,j}$ changes.}
    \label{fig:var}
\end{figure}
From \autoref{fig:var}, we see that two tasks $\tau_{i_1}$ and $\tau_{i_2}$ is scheduled on $BS_j$ from time $S$ to time $F$. The value of indicator variables $x_{i,j}(t)$, $y_{i,j}$ and $z_{i,j}$ is $0$ for both the tasks initially ($t\leq S$). For $\tau_{i_1}$ the value of $x_{i_1,j}(t)=1$ from time $S\leq t \leq F$, as it executes during that time period. So, the value of $y_{i_1,j}=1$ as it started execution and $z_{i_1,j}=0$ as it is not dropped. For $\tau_{i_2}$, the value of $x_{i_2,j}(t)=1$ only for time $S\leq t \leq D$, after that it is $0$, as $\tau_{i_2}$ is dropped at $t=D$. The value of both $y_{i_2,j}$ and $z_{i_2,j}$ is $1$, as $\tau_{i_2}$ started execution and got dropped.

We formally denote the drop penalty of all AVs experiencing service outage as $C_{drop}$, which can be calculated as, 
\begin{equation} 
\footnotesize
     C_{drop} =   \sum_{i=1}^N  \left( (1-\sum_{j=1}^M y_{i,j})+\sum_{j=1}^M z_{i,j}*y_{i,j} \right) * \left[ (1+ \delta ) * pen_i^{f} \right] 
\label{eqn:penalty}
\end{equation}
where $pen_i^f$ denotes the penalty on not completing $\tau_i$ with in its deadline. $pen_i^{f}$ can have two values between $pen^{h}$ and $pen^{s}$. The former denotes the penalty of failure for hard, and the latter denotes the penalty for missing soft real-time tasks. The ratio $\frac{pen^{h}}{pen^{s}}$ can denote how critical the hard real-time tasks are with respect to soft real-time tasks. The value of this ratio is $100:1$ as considered in \cite{TIANSEN2023}. Here $\delta$ is any value in the range $[0,1]$ decided beforehand between VECS infrastructure provider and AVs.

\textcolor{black}{
When the task $\tau_i$ is not scheduled, the value of the $(1- y_{i,j}) = 1$, and $z_{i,j} * y_{i,j} = 0$ in \autoref{eqn:penalty}, so $C_{drop}$ for that case is $(1+ \delta ) * pen_i^{f}$. If the task $\tau_i$ is scheduled and dropped, the part $(1- y_{i,j}) = 0$, and $z_{i,j} * y_{i,j} = 1$, so $C_{drop}$ for that case is $(1+ \delta ) * pen_i^{f}$. But if the task $\tau_i$ is scheduled and completed within deadline, the part $(1- y_{i,j}) = 0$, and $z_{i,j} * y_{i,j} = 0 \: (z_{i,j}=0, \: y_{i,j}=1)$, so $C_{drop}$ for that case is $0$.}

\begin{table}[]
    \centering
    \footnotesize
    \begin{tabular}{|p{2.25cm}|p{5.75cm}|}
      \hline
        \textbf{Notation} & \textbf{Definition}  \\
        \hline
        AV, BS, CS, CC & Autonomous vehicle, base station, centralized scheduler, centralized cloud \\ 
        \hline
        \textit{I}, $N$, $\tau_i$ & Set of all tasks, number of tasks, $i^{th}$ task \\
        \hline
        $\mathcal{B}$, $M$, $BS_j$ & Set of allBSs, number of BSs, $j^{th}$ BS \\
        \hline
        $id_i, a_i$, $d_i$, $s_i$, $flag_i$& Id, generation time, deadline, starting time and flag of $\tau_i$ \\
        \hline
        $pen^f_i$, $pen^h$, $pen^s$ & Penalty paid by VECS on failing to complete $\tau_i$, hard task, soft task within deadline \\
        \hline
        $AV_l$, $t_{\beta}$ & $l^{th}$ AV, length of one batch\\ 
        \hline
        $t_{b_k}$, $I_{b_k}$ & $k^{th}$ batch, set of all tasks in $k^{th batch}$\\
        \hline
        $p_i$, $p'_i$, $f_i$ & Minimum processing time, processing time using $u_i$ resource, finish time of $\tau_i$\\
        \hline
        $u^{max}$, $u_i$, $u^{min}_i$ &Maximum resource consumed by all tasks, resource consumed by $\tau_i$, minimum resource consumed by $\tau_i$\\
        \hline
        $U_j(t)$, $U^{max}$ & Utilisation consumed by tasks in $BS_j$ at time $t$, maximum proportion of total utilisation of BS used\\ 
        \hline
        $P_j(t)$, $P^{max}$, $P_S$ & Electric power usage at time $t$, maximum electric power usage of a BS, static power usage\\
        \hline
        $U(t)$, $U^T$ & Utilisation of all BS at time $t$, threshold utilisation value of TOR\\
        \hline
        $d_{i,j}$ & Distance $\tau_i$ generating AV and $BS_j$\\
        \hline
        $\delta$ & The ratio of the penalty for failing to complete a task to the task's revenue proportion\\
        \hline
        $TOR_i$, $TSR_i$ & Task Offloading Request and Task scheduling Request of $\tau_i$\\
        \hline
        $TOA_{i,s}$, $TEA_{i,s}$ & Task Offloading Approval and Task execution Approval between $\tau_i$ and $BS_s$\\
        \hline
        $LOC_{BS}$ & Vector storing location of all BS\\
        \hline
        $LOC^{t_k}_{AV}$ & Vector storing location of all AVs at $t_k$ \\
        \hline
        $x_{i,j}(t)$ & Indicator variable, 1 if $\tau_i$ is scheduled in $BS_j$ at time t, 0 otherwise \\
        \hline
        $y_{i,j}$ & Indicator variable, 1 if $\tau_i$ is scheduled in $BS_j$ at any time, 0 otherwise\\
        \hline
        $z_{i,j}$ & Indicator variable, 1 if $\tau_i$ is dropped from $BS_j$ at any time, 0 otherwise\\
        \hline
        $C_{dis}*\lambda_{dis}$, $C_{drop}$, $C_e*\lambda_e$, $C_{total}$ & Distance cost, task dropping cost, energy cost, total cost of the VECS \\
        \hline
        $\widehat{U}^{max}$&Maximum allowed power consumption of all BSs\\
        \hline
        $\widehat{D}^{max}$&Maximum allowed distance between task generating AV and task executing BS\\
        \hline
        $T_c$, $T_{total}$& Current time and total time\\
        \hline
    \end{tabular}
    \caption{Notations used}
    \label{Table_1}
\end{table}

\subsection{Problem Statement}

Given $M$ BSs and the $N$ online tasks arriving to the VECS infrastructure (of the city), our aim is to minimize the total penalty of all the BSs. $C_{total}$ denotes the overall cost penalty of the BS, and it has three components to it: drop penalty cost ($C^{drop}$), energy cost ($\lambda_e * C_{e}$) and distance cost ($\lambda_{dis} * \ C_{dis}$). Thus the total penalty is given by :

\begin{equation}
        C_{total} =  C_{drop} + \lambda_{dis} * \ C_{dis} + \ \lambda_e * C_{e}  
\end{equation} 
So, we formally write the objective of the problem in \autoref{eqn:opt}.
\begin{equation}
    min(C_{total})
    \label{eqn:opt}
\end{equation}

The constraints are as follows: 
\begin{enumerate}
    \item Each task ($\tau_i$) if assigned, should be assigned to exactly one BS. It is formally presented as $\begin{aligned}[t]
    \sum_{j=1}^M y_{i,j} \leq 1 \ \forall i\in I
    \end{aligned}$
    \item Each BS should be assigned tasks such that the sum of utilities of tasks should not exceed its maximum capacity. It is formally presented in \autoref{eqn:utility}.
    \item Each task should execute using the computation resource of BS's VM within a range in between $u^{max}$ and $u^{min}_i$. It is formally presented as $\begin{aligned}[t]
    u^{max} \geq u_{i} \geq u^{min}_{i} \ \forall i \in I
    \end{aligned}$
    
\end{enumerate}

We summarize all the symbols used in \autoref{Table_1} for ease of reading.

\section{Solution Approaches}
\textcolor{black}{The AVs generate tasks in the city, and AV's internal firmware decides whether to offload tasks to the infrastructure or not. Let us consider $AV_l$ generates $\tau_i$. $AV_l$ internal firmware decides to offload the task. If it decides, the task requires offloading then, $AV_l$ sends a $Task\ Offloading\ Request\ (TOR)$ to all available BSs with which it can establish a direct connection. We denote the $TOR$ for $\tau_i$ as $TOR_i$ and it contains metadata data of the task ($id_i$, $a_i$, $d_i$, $p_i$, $flag_i$). Upon receiving the request, each BS evaluates the task using a certain criterion and makes a decision regarding whether to accept or reject the $TOR_i$. If the $BS_s$ accepts the request, it sends $Task\ Offloading\ Approval\ (TOA)$ to $AV_l$. It is a binary signal denoting the acceptance or rejection of the request for task execution of AV. The TOA of $\tau_i$ from $BS_s$ is denoted by $TOA_{i,s}$. The BS tests whether the task can be accommodated or not with utilisation $u^{max}$, gradually moving it down to $u^{min}_i$. The BS schedules a task after receiving $TOR$ from AV only if its utilisation is less than $U^T$ with as much utilisation as possible. We denote $U^T$ as the Threshold utilisation value of TOR. On receiving $TOA_{i,s}$ from at least one BS, the $AV_l$ sends Negative Acknowledgement $NACK_i$ to the rest of the BSs. In the case, if multiple BS sends $TOA$, the AV chooses the BS whose $TOA$ reaches the AV first. On receiving $TOA_{i,s}$, the $AV_l$ offloads the task to that BS, which subsequently executes the task. The pseudo-code of AV's framework and BS framework is presented in \autoref{alg:av} and \autoref{alg:bs} respectively. When the task load is low, the VECS operates in this manner due to the reduced computational demand. Consequently, the utilization demand on the BSs is below \( U^{T} \). As a result, the requests are processed locally, a mode of operation we refer to as ``local mode".}

\begin{figure}
   
    \centering
    \includegraphics{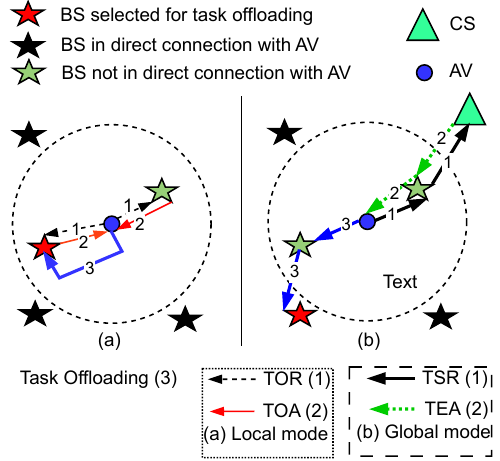}
    \caption{ \footnotesize The order of $TOR$, $TOA$, $TSR$, and $TEA$ in VECS is shown. The numbers indicate the time of each event. Figures (a) and (b) indicate local and global modes of operation, respectively.}
    \label{fig:tsrtor}
    \vspace{0.2cm}
\end{figure}

Conversely, if a task arrives at a BS, and the computation usage of that BS is more than $U^{T}$, the BS does not send the $TOA$. So, the AV waits for some time and initiates a $Task\ Scheduling\ Request\ (TSR)$ to the CS. We denote this request as $TSR_i$ for $\tau_i$ and it contains metadata data of task ($id_i$, $a_i$, $d_i$, $p_i$, $flag_i$) and location of $AV_l$. The $AV_l$ sends $TSR_i$ to the nearest BS, and the BS routes it to the CS. The CS accumulates and schedules all tasks whose $TSR$ it receives as a batch over a certain period. The CS executes \autoref{alg:cs}, and decides the BS to which the AV should offload the task and what amount of computational cores the task should execute on. On finding out the appropriate BS (let us assume it is $BS_s$), CS sends $Task\ Execution\ Approval\ (TEA)$ to $BS_s$ and $AV_l$. We denote this as $TEA_{i,s}$ as $\tau_i$ is scheduled at $BS_s$. It is a tuple consisting of BS information, AV information, computation utilisation and task execution starting batch time ($BS_s,\ AV_l,\ u_i,\ t_{b_k}$). On receiving $TEA_{i,s}$ for $\tau_i$ the AV offloads the task, while the BS allocates resources for incoming $\tau_i$. Let us assume that $\tau_i$ is scheduled at batch $t_{b_k}$ on $BS_s$. Subsequently, the $AV_l$ offloads the $\tau_i$ to $BS_s$, where the task execution starts at $t_{b_k}$. 

So, in the second case, the CS collects $TSR$ from all the BSs whose utilization is more than $U^T$ and schedules them collectively, a mode of operation we term as ``global mode". The CS is not continuously active; it is only activated when the system operates in global mode. The idea of this scheduler is inspired by \cite{worksteal}. \autoref{fig:tsrtor} shows the timing order of the entire process.

\begin{figure*}[tb!]
    \centering
    \includegraphics[width=\textwidth]{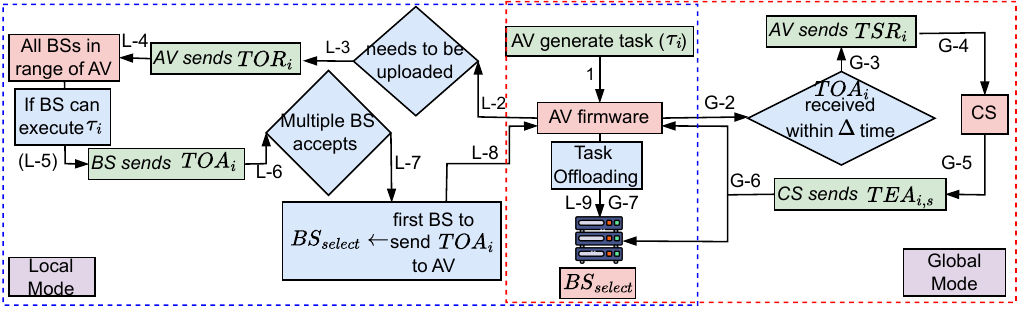}
    \caption{ \footnotesize The figure depicts the flowchart of our approach. \textcolor{myred}{\colorbox{myred}{H}} denotes a component of the VECN, \textcolor{myblue}{\colorbox{myblue}{H}} indicates a decision within the approach, \textcolor{mygreen}{\colorbox{mygreen}{H}} signifies an event of signal transmission among the VECN components, and \textcolor{myviolet}{\colorbox{myviolet}{H}} represents the two modes of scheduling discussed in the proposed approach. The sequence of events is marked by numbers. Events in local mode start with ``L" (enclosed by blue rectangle), while those in global mode (enclosed by red rectangle) start with ``G". The AV first initiates local mode execution; if this is unsuccessful, it then initiates global mode execution. If either execution mode is successful, the process concludes with task offloading to \(BS_{select}\).}
    \label{fig:fc}
\end{figure*}

We define three operations $TrySchedule$, $Schedule$ and $Execute$. The details are: (a) $TrySchedule(BS_j,\tau_i,u_i,t_{b_k})$ returns a binary value (if is it possible to schedule $\tau_i$ at $BS_j$ using $u_i$ resource for $p'_i$ processing time on $k^{th}$ batch), along with the subsequent cost ($C_{e}$ and $C_{dis}$). If it is not possible to schedule, the value of $C_e$ and $C_{dis}$ is $NULL$. (b) $Schedule(BS_j,\tau_i,u_i,t_{b_k})$ schedules $\tau_i$ at $BS_j$ using $u_i$ resource for $p'_i$ processing time on $k^{th}$ batch. It means that the computational resources of BS are reserved for the task. (c) $Execute(BS_j,\tau_i,u_i,t_{b_k})$ signifies actual usage of resources are freeing of resources after computation. \autoref{fig:fc} provides a visual representation of the system flowchart.

\begin{algorithm}
\footnotesize
    \caption{AV Framework at $AV_l$}
    \label{alg:av}
\KwIn{Generated task ($\tau_i$)}
\KwOut{$TOR_i$, $TSR_i$, task offloading}

$AV_l$ generates $\tau_i$

$AV_l$'s internal firmware decides whether to offload $\tau_i$

\For{all BS in its range}{
Send $TOR_i(id_i,a_i,d_i,p_i,flag_i)$ for respective BSs
}
\uIf{$TOA$ is received within $\Delta$ time}{
$BS_s \gets$ Choose the first BS that accepted $TOR_i$

Send $NACK_i$ to the rest of the BS
}
\uElseIf{no $TOA$ is received within $\Delta$ time}
{
Send $TSR_i(id_i,a_i,d_i,p_i,flag_i)$ to CS

Offload $\tau_i$ to the BS recommended by CS ($BS_s$)
}
\uElseIf{$AV_l$ receives $TOA_{i,s}$}
{
Offload $\tau_i$ to the $BS_s$
}
\uElseIf{$AV_l$ receives $TEA_{i,s}$ from CS}
{
Offload $\tau_i$ to the $BS_s$ as recommended by CS
}
return
\end{algorithm}

\begin{algorithm}
\footnotesize
    \caption{BS Framework at $BS_s$}
    \label{alg:bs}
\KwIn{$TOR_i$, $TSR_i$, $U(t)$}
\KwOut{$TEA_{i,s}$, task execution}

\uIf{$BS_s$ receives $TOR_i$}{
\For{$u_i$ in range $u^{max}$ to $u^{min}_i$ in descending order}{
\If{$U_j(t) + u_i \leq U^{T} \:$ for all $t \in (s_i,f_i)$ and $f_i \leq d_i$}{
Send $TOA_{i,s}$ to $AV_l$

$Scheudle(BS_s,\tau_i,u_i,t_{b_k})$; break;
}
}

}

\uElseIf{$BS_s$ receives $TSR_i$}{
Route to CS
}
\uElseIf{$BS_s$ receives $TEA_{i,s}$ from CS}{
Schedule $\tau_i$ using specified specifications in the specified time
}

\uElseIf{$BS_s$ receives $NACK$ from $AV_l$}{
Remove $TOR_i$
}

\uElseIf{$AV_l$ offloads $\tau_i$}{
Execute $\tau_i$
}

return
\end{algorithm}

\begin{algorithm}
\footnotesize
    \caption{CS Scheduling Framework for batch $t_{b_k}$}
    \label{alg:cs}
\KwIn{$I_{b_k},\:LOC_{BS},\:LOC_{AV}^{t_k},\:T_c$(Current time), $U(t)$(Utilization vector of all BSs)}
\KwOut{Updated $C_{total}$}

Set $C_{total} \leftarrow 0,\:C_{drop} \leftarrow 0,\:C_{dis} \leftarrow 0, \: C_{e} \leftarrow 0$

Set $TQ \leftarrow$ all $\tau_i$ with $d_i +p_i\leq T_c$

\While{$T_c \leq$ $T_{total}$}
{
    \For{all $\tau_i$ whose $a_i \in$ [$T_c - t_{\beta}$, $T_c$) and receives $TSR_i$ from some BS}
        {Push $\tau_i$ into $TQ$}

    \For{all $\tau_i$}
        {

        $Schedule(BS_s, \: \tau_i, \:u_i,\:t_{b_k}$) using Algo. \ref{alg:alg_selfish} or \ref{alg:alg_five} or \ref{alg:alg_dynamic} at $BS_s$

        Send $TEA_{i,s}(BS_s, \tau_i,u_i,t_{b_k})$ to $BS_s$ and $AV_l$
        }
    \For{all $\tau_i$ in TQ, expected completion time on CC $\leq d_i$}
    {
        Send the tasks to CC for execution
    }
}
Update $C_{total}$

return
\end{algorithm}

The CS follows the following strategies to schedule the tasks. These strategies are applicable only in global execution mode, as the CS is active exclusively in this mode. Additionally, in this work we consider that if the task is allocated then the task can not be migrated to any other BS or CC (but soft tasks can be dropped). 

\subsection{Task Selection Strategy}
VECS select a task based on its criticality and deadline. Like any traditional mixed critical system \cite{TIANSEN2023}, priority for task execution is given to high critical tasks. Out of the task high critical task, the CS selects a task based on EDF \cite{Yang19, Guo15} as it gives better results than other task selection policies. 

\subsection{Utilisation Selection Strategies}

In our methodology, we employ three strategies to determine the utilisation of task execution. These strategies are outlined as follows:
\begin{itemize}

\item {\textbf{Execution by maximum utilisation ($u^{max}$):} The CS selects the BS in a way that minimizes $C_{dis}+C_e$. $\tau_i$ executes for a duration $p_i$, consuming the maximum utility, $u^{max}$, of the selected BS.} 

\item{\textbf{Execution by minimum utilisation ($u^{min}$):} CS selects the BS so that $C_{dis}+C_e$ is minimum and $\tau_i$ executes for $d_i-s_i$ duration on VM with $u^{min}_i$ CPU utility of the selected BS. The CS calculates the $u^{min}_i$ using \autoref{eqn:u_min}.} 

\item{\textbf{Execution by favourable utilisation ($u^{fav}$):} CS selects utilisation in $u^{fav}$ and $\tau_i$ executes for required duration. Based on our experiments we choose VM with CPU utilisation values of $0.6,\:0.7,\:0.8$ to be in $u^{fav}$. The CS calculates the processing time using \autoref{eqn:utility}. The CS try to schedule $\tau_i$ based on descending values in $u^{fav}$.}
\end{itemize}

In all three approaches, the CS selects a BS that provides the required utilisation values based on the chosen strategy, satisfying the condition specified in \autoref{eqn:opt}. The pseudo-code for the three strategies is presented in \autoref{alg:u_fav}.

Two important terms, $\widehat{D}^{max}$ and $\widehat{U}^{max}$, are defined in our approach. $\widehat{D}^{max}$ denotes the maximum allowed distance between the AV requesting the task and the BS serving the task, while $\widehat{U}^{max}$ represents the maximum allowed power consumption in any BS. $\widehat{U}^{max}$ is a proportion of $U^{max}$. The proper values for these two parameters are analyzed through experiments, as discussed in \autoref{subsec:exp_u_d}. 

\subsection{Edge-Server Selection}
In our methodology, we select BSs in two distinct phases:
\begin{itemize}
    \item {\textbf{Selection of BS for scheduling task:} The CS selects BS where $C_e+C_{dis}$ cost is minimum. The CS evaluates all BSs where scheduling is feasible at the specified utilisation value, selecting the BS with the minimum cost. We use $BS_s$ to denote the selected BS where the combined cost of energy and distance is minimum. The pseudo-code of the process is mentioned in \autoref{alg:u_fav}.}
    \item{\textbf{Selection of BS for dropping soft tasks and adding hard tasks:} The cost associated with dropping hard tasks, $C_{drop}$, is set to be 100 times higher for hard tasks than for soft tasks. So, the CS often drops some soft tasks in order to accommodate hard tasks. The CS strategically drops soft tasks to minimize $C_{drop}$ for task dropping and reallocates the resources of the dropped soft tasks to facilitate the execution of hard tasks. We use $BS_d$ to denote the selected BS that provides the BS where dropping the cost of soft tasks is minimal. The pseudo-code of this approach is formally presented in \autoref{alg:drop}.}
\end{itemize}

\subsection{Proposed Approach}
\autoref{alg:alg_selfish}, \autoref{alg:alg_five} and \autoref{alg:alg_dynamic} shows pseudo-code for our three proposed approaches. In the proposed approach, the VECS uses dynamic holding approach \autoref{sec_dynhold} in global operation mode. We use the other two approaches for comparison purposes.

In each of these approaches, the CS selects a task from the task queue ($TQ$). Upon task selection, the CS assesses whether the task can be executed successfully within the deadline. Upon completion of task execution or upon surpassing the deadline, the CS removes the task from $TQ$. If it is possible to schedule the task within deadline or the CS drops any tasks then it updates the corresponding $Cost$ values.

\begin{algorithm}
\footnotesize
    \caption{Selection of BS and Utility}
    \label{alg:u_fav}
\KwIn{$u_k, \: LOC_{BS}, \tau_i, \: t_{b_k},\:LOC_{AV}^{t_k}$, $U(t)$}
\KwOut{$BS_s \gets$ Selected BS to execute $\tau_i$}

\For{all $BS_j$ }{
  \If{$U_j(t) + u_k \leq \widehat{U}^{max}$, $f_i \leq d_i$ and $D(Loc(BS_j),Loc(AV_l))\leq \widehat{D}^{max}$}{
    $BS_s \gets \min [(C_e+C_{dis})$ for $TrySchedule(BS_j, \tau_i , u_k,t_{b_k})]$
  }
}
return $BS_s$
\end{algorithm}

\begin{algorithm}
\footnotesize
    \caption{Task Drop}
    \label{alg:drop}
\KwIn{$u_k, \: LOC_{BS}, \tau_i, \: t_{b_k},\:LOC_{AV}^{t_k}$, $U(t)$}
\KwOut{$BS_d \gets $ BS whose soft tasks are dropped}

$BS_{d}$ $\leftarrow$ Find server which drops minimum number of soft tasks so that $Schedule(BS_d,\: \tau_i, \: u_{min}^i,\:t_{b_k})$ is successful

\If{scheduled successfully}{
$C_{total} \leftarrow C_{total} + C_{drop}$ of all the soft tasks dropped

Remove all the dropped soft tasks from $TQ$
}
return $BS_d$
\end{algorithm}

\subsubsection{Penalty Minimisation using Selfish Holding}

In this approach, the CS chooses the BS where task $\tau_i$ can execute utilizing the computational resource $u^{max}$ of that BS so that the combined distance and energy cost is minimum using \autoref{alg:u_fav}. If the scheduling attempt is unsuccessful and $\tau_i$ is a hard task, the CS seeks a BS by dropping soft tasks using \autoref{alg:drop} where $C_{drop}$ is minimum and the CS reallocates the resources of the dropped tasks to facilitate a VM for execution of the hard task. The entire approach is detailed in \autoref{alg:alg_selfish}.
\begin{algorithm}
\footnotesize
    \caption{Proposed Selfish Holding Approach}
    \label{alg:alg_selfish}
\KwIn{$LOC_{BS}, \tau_i, \: t_{b_k},\:LOC_{AV}^{t_k}$, $U(t)$}
\KwOut{Scheduled tasks}
Sort the $TQ$ based on $flag^h$ and deadline

\For{each $\tau_i$ in the $TQ$}{
\eIf{$\tau_i(d_i)$ not passed}
{
$BS_s \gets$\autoref{alg:u_fav}($u^{max}, \: LOC_{BS}, \tau_i, \: t_{b_k},\:LOC_{AV}^{t_k},\:U(t)$)

\eIf{$TrySchedule(BS_s,\: \tau_i, \: u^{max},\:t_{b_k})$}{
$Schedule(BS_s,\:\tau_i,\:u^{max},t_{b_k})$

Remove $\tau_i$ from $TQ$, update $C_{dis},\:C_{e}$, continue

}{
\If{$\tau_i(flag)==h$}{
$BS_d \gets$\autoref{alg:drop} ($u^{min}_i, \: LOC_{BS}, \tau_i, \: t_{b_k},\:LOC_{AV}^{t_k},\:U(t)$)

$Schedule(BS_d,\:\tau_i,\:u^{min}_i,t_{b_k})$

Remove $\tau_i$ from $TQ$, update $C_{drop},\:C_{dis},\:C_{e}$
}
}
}
{Remove $\tau_i$ from $TQ$, update $C_{drop}$}
}
\end{algorithm}

\subsubsection{Scheduling to the Nearest Server}

In contrast to the previous approach, the CS in this strategy selects a BS where $C_{dis}$ is minimized. The CS executes $\tau_i$ utilizing the computational resource $u_i^{min}$ of that BS. Similar to the previous approach, the CS drops soft tasks using \autoref{alg:drop} and employs the resources of the dropped tasks to facilitate the execution of hard tasks. We use $BS_n$ to represent the BS nearest to the location of AV generating the task. The overall approach is represented in \autoref{alg:alg_five}.
\begin{algorithm}
\footnotesize
    \caption{Nearest Scheduling Approach}
    \label{alg:alg_five}
\KwIn{$LOC_{BS}, \tau_i, \: t_{b_k},\:LOC_{AV}^{t_k}$, $U(t)$}
\KwOut{Scheduled tasks}
Sort the $TQ$ based on $flag^h$ and deadline

\For{each $\tau_i$ in the $TQ$}{
\eIf{$\tau_i(d_i)$ not passed}
{
$BS_n \gets$ Nearest Server to $AV_l$

\eIf{$TrySchedule(BS_n,\: \tau_i, \: u^{min}_i,\:t_{b_k})$}{
$Schedule(BS_n,\: \tau_i, \: u^{min}_i,\:t_{b_k})$

Remove $\tau_i$ from $TQ$, update $C_{dis},\:C_{e}$, continue
}
{
\If{$\tau_i(flag)==h$}{
$BS_d \gets$\autoref{alg:drop}($u_i^{min}, LOC_{BS}, \tau_i, \: t_{b_k},$

$\:LOC_{AV}^{t_k},\:U(t)$)

$Schedule(BS_d,\:\tau_i,\:u_i^{min},\:t_{b_k})$

Remove $\tau_i$ from $TQ$, update $C_{drop},\:C_{dis},\:C_{e}$
}
}
}
{Remove $\tau_i$ from $TQ$, update $C_{drop}$}
}
\end{algorithm}

\subsubsection{Penalty Minimisation using Dynamic Holding}
\label{sec_dynhold}

In this particular strategy, the CS a specialized approach to enhance performance. 
In this approach, the CS selects BS where $\tau_i$ utilizes computational resources within the range of $u^{fav}$. If it is not feasible to find a BS that meets the computational requirements in that range, the CS selects a BS using the $\tau_i$ executes using minimum computational resource $u^{min}_i$. If $Schedule$ is still not successful and and the task is categorized as hard, the CS uses \autoref{alg:drop} to find BS and replace soft tasks with hard tasks reusing the computation resource. The pseudo-code of the entire approach is presented in \autoref{alg:alg_dynamic}.

\begin{algorithm}
\footnotesize
    \caption{Proposed Dynamic Holding Approach}
    \label{alg:alg_dynamic}
\KwIn{$LOC_{BS}, \tau_i, \: t_{b_k},\:LOC_{AV}^{t_k}$, $U(t)$}
\KwOut{Scheduled tasks}
Sort the $TQ$ based on $flag^h$ and deadline

\For{each $\tau_i$ in the $TQ$ \label{l2}} {
\eIf{$\tau_i(d_i)$ not passed}
{

\For{$u_k$ in $u^{fav}$ in descending order}{

$BS_s \gets$ \autoref{alg:u_fav}($u_k, \: LOC_{BS}, \tau_i, \: t_{b_k},\:LOC_{AV}^{t_k},\:U(t)$)

\If{$TrySchedule(BS_s,\: \tau_i, \: u_k,\:t_{b_k})$}{
$Schedule(BS_s,\: \tau_i, \: u_k,\:t_{b_k})$

Remove $\tau_i$ from $TQ$, update $C_{dis},\:C_{e}$, continue

}
}

\eIf{$TrySchedule(BS_s, \: \tau_i, \: u_{min}^i,\:t_{b_k})$}{
$BS_s \gets$ \autoref{alg:u_fav}($u_{min}^i, \: LOC_{BS}, \tau_i, \: t_{b_k},\:LOC_{AV}^{t_k},\:U(t)$)

$Schedule(BS_s, \: \tau_i, \: u_{min}^i,\:t_{b_k})$

Remove $\tau_i$ from $TQ$, update $C_{dis},\:C_{e}$, continue
}{
\If{$\tau_i(flag)==h$}{
$BS_d \gets$\autoref{alg:drop}($u_{min}^i, \: LOC_{BS}, \tau_i,$

$t_{b_k},\:LOC_{AV}^{t_k},\:I_{t_{b_{k n}}}$)

$Schedule(BS_d,\:\tau_i,\:u^i_{min},\:t_{b_k})$

Remove $\tau_i$ from $TQ$, update $C_{drop},\:C_{dis},\:C_{e}$
}

}

}
{Remove $\tau_i$ from $TQ$, update $C_{drop}$}
}
\end{algorithm}

\section{Experimental Results}
\subsection{Dataset}
\subsubsection{Synthetic Dataset}

For any task, arrival time is generated randomly between [0,2000] and batch time is chosen as $t_{\beta}=3$ unit, the deadline is generated based on the above constraint. The processing time is generated randomly such that it lies in between the arrival time and the deadline. The flag for the hard or soft task is generated randomly as $flag_i=\{h,s\}$. For example for any task $\tau_i$, arrival time is determined by $a_i = rand(0, 2000)$. Let $\tau_i$ falls in the time slot [$t{b_k}$,$t_{b_k}+t_{\beta}$]. The deadline is determined by $d_i \geq t{b_k} + t_{\beta}$. The processing time $p_i$ is determined by $(0,1)*(d_i-a_i)$. The total number of tasks is 1 million.

\subsubsection{Real Life Trace Data}
\label{lab_real_fata}

Since the considered problem is new, we can not find an exact real life dataset. Instead we merged two dataset to create the required data for this model. We extracted the locations of BSs and AVs from the dataset provided in Lai et al. \cite{locdata}. We extracted task data, i.e. arrival time, execution time, deadline, etc. from the dataset provided in Wilkes et al. \cite{clusterdata:Wilkes2011, clusterdata:Reiss2011}.

The locations are given as longitude and latitudes in the location data, and we converted them to planar coordinates. Then, we scaled these planar coordinates to integers in the range $[0,m]$. Similarly, processing time, deadline and arrival time are considered in nanoseconds in the task data, and we scaled it to integers in the range $[0,T_{max}]$. Then we choose N tasks and M BSs from Wilkes et al. and map them with locations from Lai et al. Also, the value of $U^{T}$ is fixed to $0.5*U^{max}$ and the value of $\delta$ is fixed to $0.3$.

\subsection{Slack and Ratio of hard and soft tasks}
To comprehensively compare our approach with other methodologies across various conditions, we introduce two additional terms:

\begin{itemize}
    \item {We define \textit{slack} by symbol $\mu$. We define the value of $\mu = \frac{d_i - a_i}{p_i}$. When the value of $\mu$ is greater than 3, tasks are in a loose range, and the deadline is relatively far off compared to processing, thus giving more flexibility to complete the task. When the value of $\mu$ is between 1.5 and 3, tasks are in the normal range. The deadline and processing time are relatively balanced, and the tasks should be completed in a reasonable time. When the value of $\mu$ is less than 1.5, tasks are in a tight range of deadline and processing time and the task should be completed as early as possible.}
    \item {We define another parameter $\rho$ which is the ratio between number of hard and soft tasks in a batch, which is $\rho =$ the number of hard tasks to the number of soft tasks.}
\end{itemize}

\subsection{State of the Art Approach}
To compare the results of our approach, we use the Baruah approach \cite{baruah}. In Baruah, the problem is addressed by executing tasks at maximum utilization, assuming that tasks are periodic, which differs from our scenario. In Szalay et al. \cite{rt-faas}, the same problem is tackled with mixed critical and periodic tasks. We adapt their approach to suit our case, where tasks are aperiodic and mixed critical. Modifications are necessary to align their methodology with our problem context. For comparison, we reinterpret Baruah’s objective function and assign tasks to the processor with minimum utilization.

\subsection{Evaluation of $\widehat{U}^{max}$ and $\widehat{D}^{max}$}
\label{subsec:exp_u_d}
Before we evaluate our proposed approach, we conduct experiments to choose the hyperparameters $\widehat{U}^{max}$ and $\widehat{D}^{max}$. The term $\widehat{U}^{max}$ denotes the maximum allowed of power consumption by any BS. It is a proportion of $U^{max}$. $\widehat{D}^{max}$ signifies the maximum allowed distance between the location of the BS providing the service and the location of the AV that generated the task.  

\begin{figure}[tb!]
\centering
\includegraphics[scale=.5]{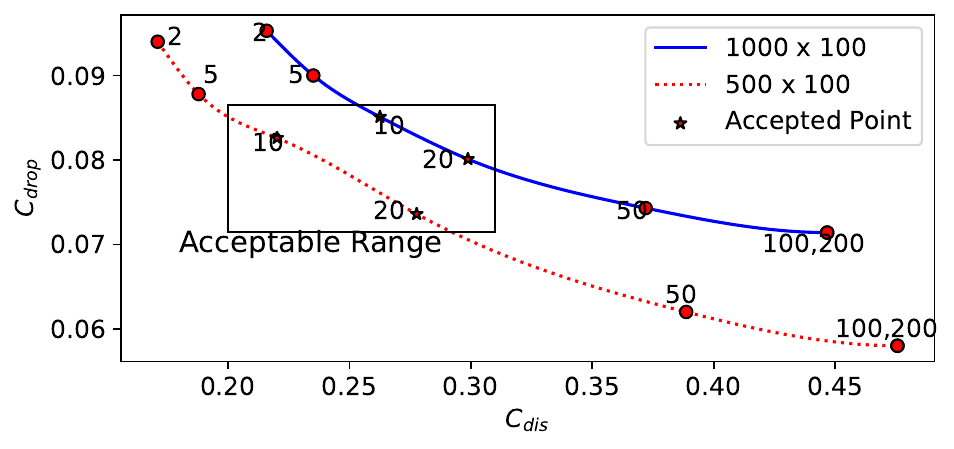}
\caption{\footnotesize Variation of $C_{drop}$ v/s $C_{dis}$ for different datasets by fixing $\widehat{U}^{max}$ and $\widehat{D}^{max}$ varying..}
\label{fig:drop_vs_dis_hyp}
\end{figure}

\begin{figure}[tb!]
\centering
\includegraphics[scale=.5]{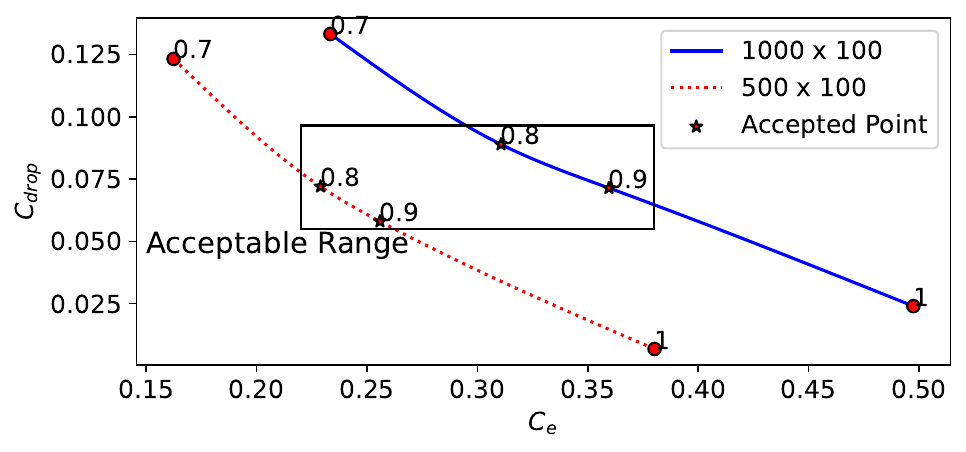}
\caption{\footnotesize Variation of $C_{drop}$ v/s $C_e$ for different datasets by fixing $\widehat{D}^{max}$ and $\widehat{U}^{max}$ varying.}
\label{fig:drop_vs_energy_hyp}
\end{figure}

We plot the variation between $C_{drop}$ and $C_{dis}$, $C_{drop}$ and $C_{e}$ for two datasets in \autoref{fig:drop_vs_dis_hyp} and \autoref{fig:drop_vs_energy_hyp} respectively. 
$N \times M$ signifies the number of tasks considered and the number of BS considered in labels in the above mentioned figures. We consider two datasets where $N \times M = 1000 \times 100$ and $500 \times 100$. The $C_{drop}$, $C_e$ and $C_{dis}$ are normalised with respect to the total profit in $I$.

In \autoref{fig:drop_vs_dis_hyp}, we first fix $\widehat{U}^{max}$ and vary $\widehat{D}^{max} \in \{2, 5, 10,20, 50, 100, 200\}$ and plot the $C_{drop}$ v\/s $C_{dis}$. Now we observe that if $\widehat{D}^{max}$ is very less, then $C_{dis}$ is very less, but $C_{drop}$ is very high as more number of soft tasks are dropped. If $\widehat{D}^{max}$ is very high, then $C^{drop}$ is very low as most of the soft tasks are also completed, but $C_{dis}$ is very high. Hence, this is a trade-off between $C_{dis}$ and $C_{drop}$. $C_{total}$ consists of both, so we want both values as small as possible and that occur in the middle range where $\widehat{D}^{max} \in \{ 10, 20\}$. So, we choose $\widehat{D}^{max} \in \{10,20\}$ for our proposed heuristic approach.  

Similarly, in \autoref{fig:drop_vs_energy_hyp} we fix $\widehat{D}^{max}$ and vary $\widehat{U}^{max} \in \{0.7, 0.8, 0.9, 1\}$ and plot the $C_{drop}$ Vs $C_{e}$.  We fixed $P_S$ and $P^{max}$ as 0.2 and 2000 units as per \autoref{eqn:energy}. The $C_{drop}$ and $C_{e}$ are normalised with respect to the total profit in $I$. We observe that if $\widehat{U}^{max}$ is less, then $C_{e}$ is less but $C_{drop}$ is very high as BS is not able to complete more tasks, thus more tasks are dropped. Similarly, if $\widehat{U}^{max}$ is very high, then $C_{drop}$ is very low as more tasks are completed, but $C_e$ is very high.

Hence, this is a trade-off between $C_e$ and $C_{drop}$. Our $C_{total}$ has both values, so we want both values to be as small as possible, and that occurs when $\widehat{U}^{max}$ is in the range of $0,8, \ 0.9$. Now, we consider a total of 4 variations of our proposed heuristic solution with $\widehat{D}^{max}$ as 10 and 20 and $\widehat{U}^{max}$ as 0.8 and 0.9.

\subsection{Analysis using Synthetic Dataset}
In this subsection, we do not consider the $C_{dis}$ and $C_e$. Here we plot the variation of $C_{drop}$ in different conditions of incoming tasks.

\subsubsection{Variation in Number of Tasks and Processors}
\begin{figure}
    \centering
    \includegraphics[scale=.5]{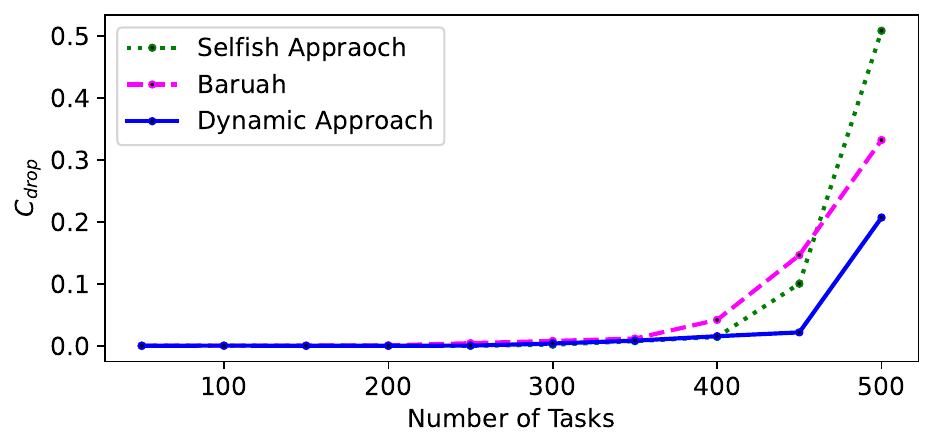}
    \caption{\footnotesize Variation of number of Tasks, $M$ = 50, $\mu$ = ``Normal", $\rho=2:1$}
    \label{fig:task_p1}
\end{figure}

\begin{figure}
    \centering
    \includegraphics[scale=.5]{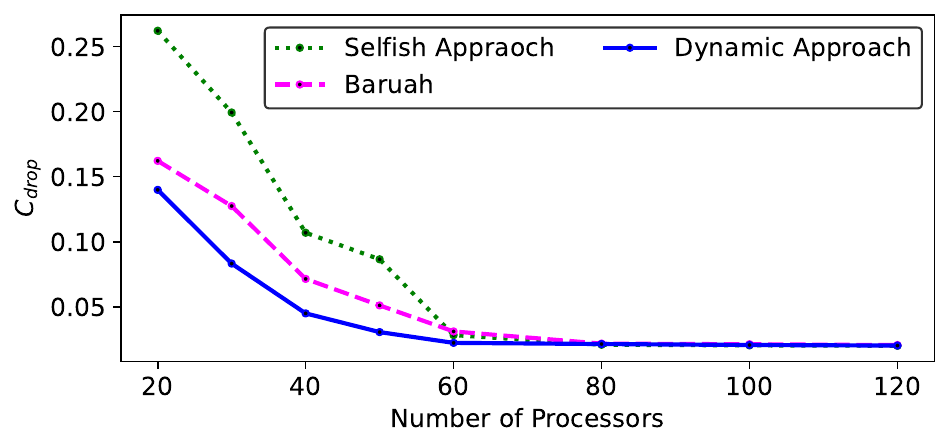}
    \caption{\footnotesize Variation of number of BSs, $N$ = 500, $\mu$ = ``Normal", $\rho=1:1$}
    \label{fig:proc_p1}
\end{figure}
 \autoref{fig:task_p1} shows the effect of variation number of tasks ($N$) on $C_{drop}$, keeping the number of BSs ($M$) fixed to 50. We observe that initially, when the number of tasks is less, all approaches give similar $C_{drop}$ as most of the tasks are scheduled and completed. This is because 50 BSs are more than capable enough to provide computation for 350 or less tasks. But as the number of tasks increases, after one point, the dropping of hard tasks is inevitable, so the penalty starts increasing rapidly. 

\autoref{fig:proc_p1} shows the effect of the variation of the number of BSs ($M$) on $C_{drop}$, keeping the number of tasks ($N$) fixed to 500. We observe that as the BSs increase, more tasks can be scheduled, so $C_{drop}$ starts decreasing. After some increase when all the tasks can be scheduled, increasing the number of BSs does not affect the penalty. This is because the number of BSs is capable enough to handle the required number of tasks. The proposed dynamic utilisation approach has the least penalty in both high and low amounts of processors and tasks.  

\subsubsection{Variation in Slack of Tasks}

\begin{figure}
    \centering
    \includegraphics[scale=.5]{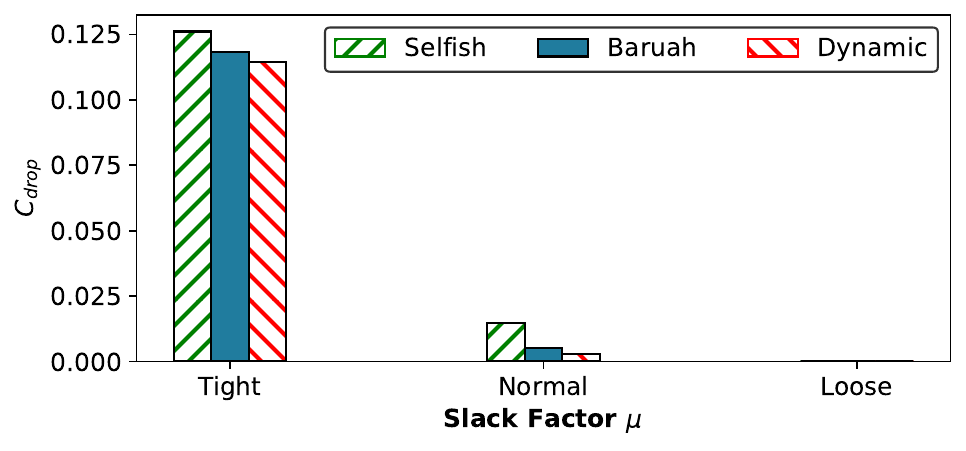}
    \caption{\footnotesize Effect on $C_{drop}$ with variation of ``slack" of deadlines, $N=500$, $M=50$, $\rho=1:1$}
    \label{fig:tight_p1}
\end{figure}

\autoref{fig:tight_p1} shows the effect of the variation of ``slack" of deadlines on $C_{drop}$. We observed that if the slack of tasks is tighter since most tasks have very short deadlines and high processing time, we cannot schedule all the hard tasks. Hence $C_{drop}$ is very high. For normal and loose slack, most tasks are scheduled and completed, hence $C_{drop}$ is low. The proposed dynamic utilisation approach has the least $C_{drop}$ for all types of slacks.

\subsubsection{Variation in Proportion of Hard Tasks}

\begin{figure}
    \centering
    \includegraphics[scale=.5]{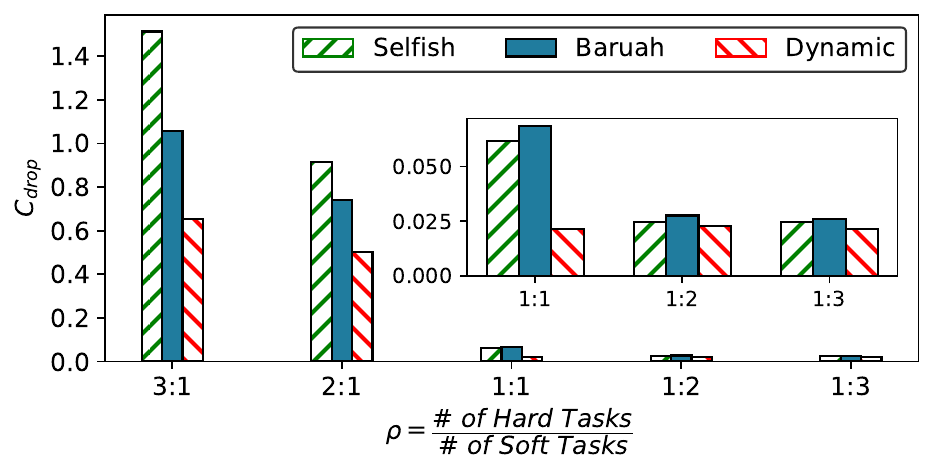}
    \caption{\footnotesize Effect on $C_{drop}$ with variation in the proportion of hard and soft tasks, $N=500$, $M=50$, $\mu$=``Normal"}
    \label{fig:hard_p1}
\end{figure}

 \autoref{fig:hard_p1} shows the effect of variation in the proportion of hard and soft tasks on $C_{drop}$. If the proportion of hard tasks increases in a batch, $C_{drop}$ increases. As the number of hard tasks decreases, the $C_{drop}$ also reduces. This is because the penalty associated with not completing hard tasks within the deadline is higher than soft tasks (as $pen^h:pen^s = 100:1$).  The proposed dynamic utilisation approach has the least $C_{drop}$ in all proportions of hard and soft tasks. 

\subsection{Analysis on Real Life Dataset}
In this subsection, we plot the variation of $C_{total}$ in different conditions of incoming tasks. The details of real-life trace data are given in \autoref{lab_real_fata}. We check the results for 4 variations of our proposed approaches: (a) $\widehat{U}^{max}=0.8,\:\widehat{D}^{max}=10$, (b) $\widehat{U}^{max}=0.9,\:\widehat{D}^{max}=10$, (c) $\widehat{U}^{max}=0.8,\:\widehat{D}^{max}=20$, (d) $\widehat{U}^{max}=0.9,\:\widehat{D}^{max}=20$. In the labels we present them as $Proposed(\widehat{U}^{max},\widehat{D}^{max})$ in \autoref{fig:diff_p2}, \autoref{fig:tight_p2}, \autoref{fig:hard_p2}.
\subsubsection{Variation in Number of Tasks and Processors}

\begin{figure}
    \centering
    \includegraphics[scale=.5]{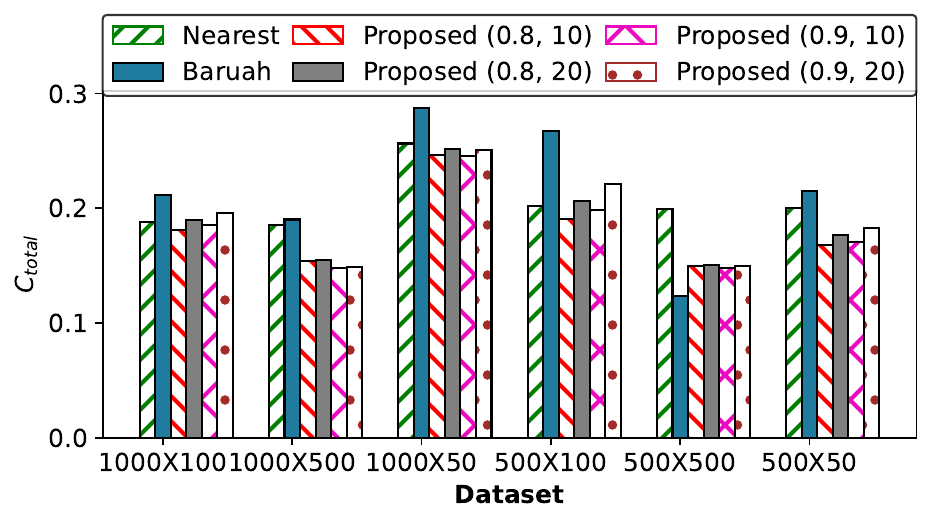}
    \caption{\footnotesize Variation in the number of tasks and BSs in real world traces, $\mu$ = ``Normal", $\rho=1:1$ }
    \label{fig:diff_p2}
\end{figure}
\autoref{fig:diff_p2} shows the performance of different approaches with our proposed approaches in different combinations of BSs and tasks. We see that $C_{total}$ decreases if the number of tasks per BS decreases. We see that our proposed method outperformed Baruah's approach in all the cases. The proposed approach outperforms Baruah's approach by 25\% when there are 500 tasks and 100 AVs.

\subsubsection{Variation in Slack of Tasks}
\begin{figure}
    \centering
    \includegraphics[scale=.5]{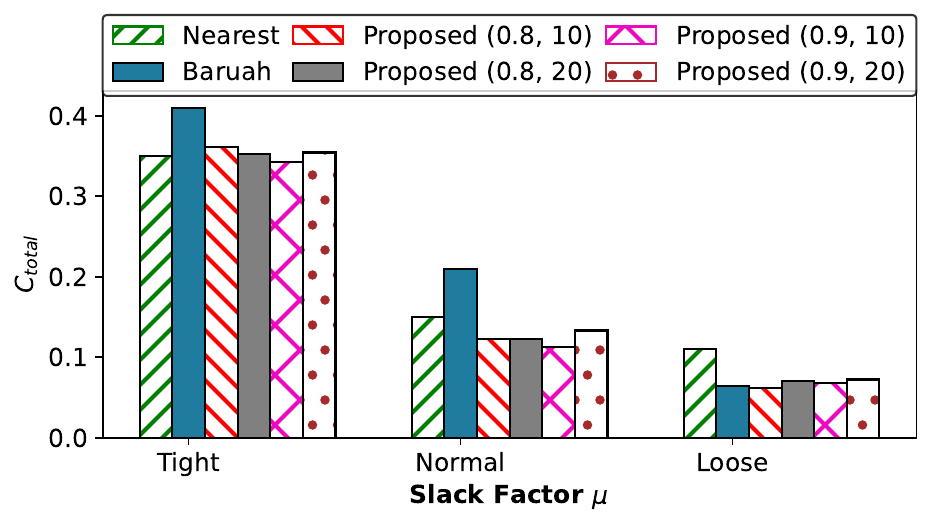}
    \caption{\footnotesize Variation in ``slack" of tasks in real world trace data, $N=1000$, $M=100$, $\rho=1:1$}
    \label{fig:tight_p2}
\end{figure}

\autoref{fig:tight_p2}  shows the comparison of the performance of different approaches with our proposed approaches in different slack of incoming tasks. We see that $C_{total}$ decreases, as the $slack$ of tasks decrease, like we see for synthetic dataset. We see that our proposed method outperformed Baruah's approach in the case when the situation is \textit{tight} and \textit{normal}. In the situation of \textit{loose} slack, all methods perform likewise. 

\subsubsection{Variation in Proportion of Hard Tasks}

\begin{figure}
    \centering
    \includegraphics[scale=.5]{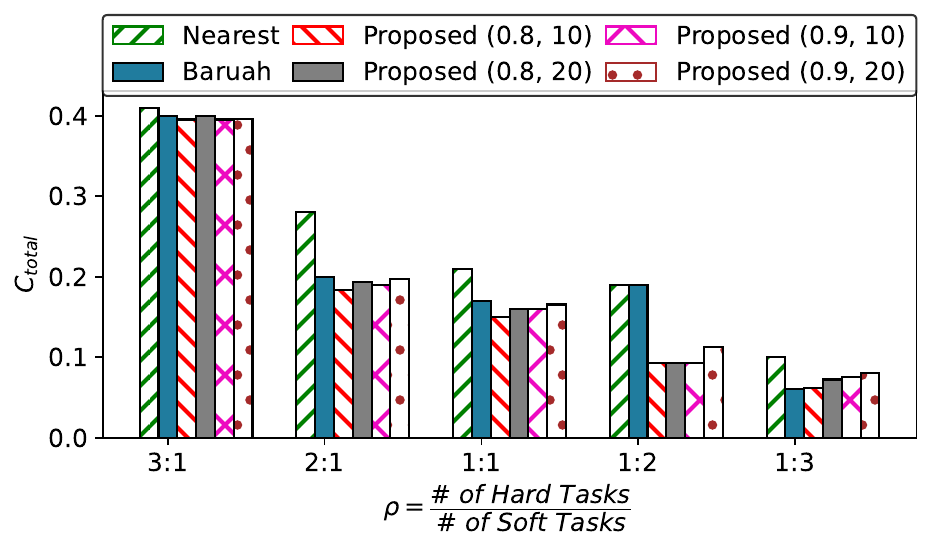}
    \caption{\footnotesize Variation in the proportion of hard tasks in real world trace data, $N=1000$, $M=100$, $\mu$ = ``Normal"}
    \label{fig:hard_p2}
\end{figure}
In \autoref{fig:hard_p2} we compare the performance of different approaches with our proposed approaches in different proportions of hard and soft tasks. We see that $C_{total}$ decreases as the number proportion of hard tasks decreases, as we see for the synthetic dataset. We see that our proposed method outperformed Baruah's approach in all the proportions.

We see that our proposed methods, give almost similar results in all the variations. Out of all our proposed methods, the method with $\widehat{D}^{max}=20$ and $\widehat{U}^{max}=0.9$ gives the best result in most of the cases.   

\section{Conclusion and Future Work}
The scheduling of mixed-critical tasks in the vehicular domain is a very interesting and challenging problem. Our proposed approach has outperformed state-of-the-art solutions on real-life datasets by 25 \%. However, there are modifications we plan to integrate into our future work. We considered the base stations' computational units to be homogeneous, and we plan to extend this to a heterogeneous system. Additionally, we aim to extend the problem to include more than two priority labels, making it more general. Addressing these issues and exploring them further could provide valuable insights for real-world applications.

\nocite{GhoseOnline}

\setstretch{0.8}
\footnotesize

\end{document}